\documentclass[twocolumn,showpacs,preprintnumbers,amsmath,amssymb,aps]{revtex4}

\usepackage{graphicx}
\usepackage{amssymb}
\usepackage{epstopdf}
\usepackage{bm}

\begin{document}
    
\title{What About Quantum Theory? Bayes and the Born Interpretation}
    
\author{F.~J.~Tipler}
\affiliation{Department of Mathematics and Department of Physics, Tulane University, New Orleans, LA 70118}

\date{\today}

\begin{abstract}
I show that probabilities in quantum mechanics are a measure of belief in the presence of human ignorance, just like all other probabilities.  The Born interpretation of the square of modulus of the wave function arises from the interaction of a quantum system with an observer, and probabilities will not arise unless the observer is unable to access all the information available in the system's wave function.  Quantum mechanics generally does not permit all the information to be obtained, even in principle, just as in relativity information from outside the past light cone cannot be obtained.  But probabilities do {\it not} imply indeterminism.  Instead, quantum mechanics is {\it more} deterministic than classical mechanics.
\end{abstract}

\maketitle

There has been a great renewal of interest in the Baynesian approach to probability theory among physicists in recent years, for examples \cite{Jaynes03},\cite{Sivia96}.  Bayesian probability theory claims that a probability is not a relative frequency, but rather a precise measure of human ignorance.  However, the use of probability in quantum mechanics has been a problem for the Bayesians ever since Born introduced his interpretation of the wave function in the 1920's.  According to Born, the square of the modulus of the wave function is a probability density, in the frequency sense of ``probability."  Since von Neumann showed that no ``hidden variable'' theory could ever reproduce the observations of quantum mechanics, this would seem to settle the debate on the meaning of probability in favor of the frequency theorists.  

The Bayesian D. S. Sivia says however, ``At this juncture, some might argue that $\ldots$ {\it chaotic} and {\it quantum} systems privide examples of physical situations which are intrinsically random.  In fact, chaos theory underlies the point we are trying to make: the apparent randomness in the long-term behaviour of a classical system arises because we do not, or cannot, know its initial conditions well enough: the actual temporal evolution is entirely deterministic, and obeys Newton's Second Law of motion.  The quantum case is more difficult to address $\cdots$  The sub-atomic world notwithstanding, it seems that `randomness' represents our inability to predict things which, in turn, reflects our lack of knowledge about the system of interest.  This is again consistent with the Bayes and Laplace view of probability, rather than the asserted physical objectivity of the frequentist approach \cite{Sivia96}, pp. 8--9.''

The Bayesian Edward Jaynes, in a entire section entitled ``But What about Quantum Theory?'' in his seminal book {\it Probability Theory: The Logic of Science} makes the same point as Sivia:  ``Those who cling to a belief in the existence of `physical probabilities'  $\ldots$ [point] to quantum theory, in which physical probabilities appear to express the most fundamental laws of physics.  $\ldots$ in quantum theory today $\ldots$, when no cause is apparent, one simply postulates that no cause exists Ñ ergo, the laws of physics are indeterministic and can be expressed only in probability form.  $\ldots$  Quantum physicsts have only probability laws because for two generations we have been indoctrinated not to believe in causes --- and so we have stopped looking for them \cite{Jaynes03}, pp. 327--328.''

I have taken the title of this paper from the title of Jaynes' section, because I shall show that we don't need to search for anything beyond quantum mechanics --- von Neumann was right --- but Jaynes and Sivia are also right: probability, even in quantum mechanics is, as in all other fields, a measure of human ignorance.

However, Jaynes overlooked a key fact about quantum mechanics: the time evolution of the wave function is completely deterministic.  Max Planck emphasized this crucial point ``Causality in Nature":  ``It is an essential fact, however, that the magnitude which is characteristic for the material waves is the wave function, by means of which the initial conditions and the final conditions are completely determined for all times and places $\dots$ We see then that there is fully as rigid a determinism in the world image of quantum mechanics as in that of classical physics.'' \cite{Planck}, pp. 64--65.

The key idea of this paper is that the square of the wave function measures, not a probability density, but a density of universes in the multiverse. If the deterministic evolution of the wave function is to mean physically real determinism, it can have no other meaning than the time evolution of the multiverse.  It is well-known that if the quantum formalism applies to all reality, both to atoms, to humans, to planets and to the universe itself, then the Many-Worlds Interpretation is trivially true (to use an expression of Stephan Hawking, expressed to me in a private conversation).  I shall show probabilities arise in quantum mechanics as they do elsewhere, as measures of human ignorance, in this case as ignorance of the other universes of the multiverse.  The probabilities arise because of the existence of the analogues of the experimenters in the multiverse, or more precisely, because before the measurements are carried out, the analogues are ``indistinguishable'' in the quantum mechanical sense.  Indistinguishability of the analogues of a single human observer means that  the standard group transformation argument used in Bayesian theory to assign probabilities can be applied.  I show that the group transformation argument yields probabilities in the Bayesian sense, and that in the limit of an infinite number of measurements, the relative frequencies must approach these probabilities.  Which is why physicists have been able to get away with assuming, for two generations, that probabilities are relative frequencies.

What the measurements of the relative frequencies actually establish is the real existence of the human analogues of the experiment, one human analogue for each possible frequency that can be obtained.  Finally physicists can give up their search for a hidden variable theory.  Quantum mechanics is its own hidden variable theory, with the other hidden universes of the multiverse being the hidden variables.  This hidden variable theory is completely deterministic, as required.

Physicists have always been very unhappy with the idea that there is an intrinsic randomness in nature.  They have always felt in their heart of hearts that the ultimate theory of reality, the long-sought Theory of Everything, ought to be completely deterministic. 

Einstein expressed the natural physicist's hostility to ultimate indeterminism very eloquently: ``That nonsense is not merely nonsense.  It is objectionable nonsense.  $\ldots$ Indeterminism is quite an illogical concept. $\dots$ The indeterminism which belongs to quantum mechanics is a subjective indeterminism.  It must be related to something, else indeterminism has no meaning, and here it is related to our own inability to follow the course of individual atoms and forecast their activities.  To say that the arrival of train in Berlin is indetermined is to talk nonsense unless you say in regard to what it it is indetermined.  If it arrives at all it is determined by something.  And the same is true of the course of atoms \cite{EinsteinPlanck}.''

I shall show that Einstein was correct.  Probability in quantum mechanics is indeed a subjective concept, related to our inability to follow the physics of individual atoms.  We cannot follow the physics of individual atoms because the behavior of individual atoms is determined by what is happening in the other universes of the multiverse, and being unaware of these other universes, we naturally cannot keep track of what is happening in them. This is the origin of probability in quantum mechanics, as I shall show using the Bohm-Landau picture.

In the opening paragrah of his ``Essay on Probabilities,'' Laplace wrote: ``All events, even those that on account of their insignificance do not seem to follow the great laws of Nature, are a result of it just as necessarily as the revolutions of the Sun.  In ignorance of the ties which unite such events to the entire system of the universe, they have been made to depend on final causes, or on chance, according as they occur and are repeated with regularity, or appear without regard to order; but these imaginary causes have gradually receded with the widening bounds of knowledge, and disappear entirely before sound philosophy, which sees in them only the expression of our ignorance of the true causes.'' (\cite{Laplace}, p. 3).

Just so.  Probabilities are a numerical measure of human ignorance, and not an objective feature of Nature herself.  In the case of quantum mechanics, the ignorance in question is ignorance of the other universes of the multiverse.  The idea that ``probability'' is a numerical measure of human ignorance was accepted by Gauss, Cauchy, Poisson, Maxwell, Boltzmann, and Gibbs, but was rejected in favor of the frequency theory of probability invented by social scientists and evolutionary biologists ignorant of physics, as has been demonstrated by the late physicist Edward Jaynes (\cite{Jaynes03}, pp. 315--317).  Since the original creators of the frequency interpretation of probability were followers of Charles Darwin, it is an interesting question whether they were motivated to discover an intrinsic fundamental randomness in nature.  

In the last chapter of his book {\it The Variation of Animals and Plants Under Domestication}, Darwin wrote:``É if we assume that each particular variation was from the beginning of all time preordained, the plasticity of organization, which leads to many injurious deviations of structure, as well as that redundant power of reproduction which inevitably leads to a struggle for existence, and as a consequence, to the natural selection or survival of the fittest, must appear to us superfluous laws of nature.'' (quote on page 432 of Volume 2 of the first edition of {\it The Variation of Animals and Plants Under Domestication} \cite{Darwin1868}, a copy of which I own).  In other words, Darwinism would be more compelling if the variations were truly random, fundamentally undetermined.  In his more widely known but earlier {\it On the Origin of Species}, Darwin had written ``I have hitherto sometimes spoken as if the variations Ñ so common and multiform in organic beings under domestication, and in a lesser degree in those in a state of nature Ñ had been due to chance.  This, of course, is a wholly incorrect expression, but it serves to acknowledge plainly our ignorance of the cause of each particular variation.'' (quote on page 131 of the second edition of \cite{Darwin1860}, a copy of which I own.)  

Darwin's remark on the superfluous nature of the natural selection law in his {\it Variations} book, written some 8 years later his {\it Origins} book, is a more accurate portrayal of the actual intellectual position of Darwinism in a deterministic cosmos, and it would not be surprising if his later followers also realized this.  However, if the ultimate physical laws had some intrinsic randomness built into them, Darwinism would be an ultimate theory of life.  Whereas if the universe were totally deterministic, living organisms no less than atoms would have their actions and offspring completely determined in the beginning of time, and biologists could expect to be eventually replaced by physicists who would be able to calculate which future species would evolve and when they would evolve.

Whatever the original source of the frequency interpretation, when the Born Interpretation was first presented in the late 1920's, the original physicists' meaning of ``probability'' had been forgotten, even by the new generation of physicists, and the frequency interpretation of probability adopted to giving a meaning to the square of the modulus of the wave function\cite{Schroedinger}.  Arnold Sommerfeld, in his 1930 book {\it Wave Mechanics} \cite{Sommerfeld}, the first textbook on the final version of non-relativistic quantum mechanics (the version of quantum mechanics taught to students today), shows no awareness that an alternative to the frequency interpretation of probability ever existed.  Nor was any awareness of the meaning given to ``probability'' by Laplace, by Poission, by Gauss, by Cauchy, or by Poincar\'e shown by the authors of any of the references to the Born Interpretation cited by Sommerfeld: \cite{Born}, \cite{Jordon}, \cite{Heisenberg}, \cite{Dirac}, \cite{London}, and \cite{BHJ}.  These authors were cited on page 83 of the first English translation, and I own C.J. Davisson's copy of this translation.  The two most recent histories on the history of probability theory, \cite{Plato} and \cite{Howie}, do not even address the question of why the probability theory developed by the leading theoretical physicists of the early 19th century was completely unknown to the leading theoretical physicists of the early 20th century.  These histories do point out that the word ``Bayesian'' was not the standard term for the probability theory developed by the above great physicists until after the Second World War.

 It is unfortunate that mathematicians have followed the physicists in thinking the frequency interpretation as fundamental, and accepted the Kolmogorov axioms as the foundation of this interpretation.  As emphasized by Jaynes (\cite{Jaynes03}, pp. 651--655), there is nothing mathematically wrong with the Kolmogorov Axioms; the are consistent, concise, and probably the best possible set of axioms for the frequency interpretation.  But they are too limited a set of axioms: they place too strong a limit on the problems that can be attacked by probability theory.  By basing probability theory on the Kolmogorov axioms, mathematics departments the world over did the equivalent of voting to declare $\pi$, the ratio of a circleÕs circumference to its diameter, equal to exactly three.  Professional mathematicians, in other words, did precisely what they ridicule state legislatures of considering:  determine a matter of mathematical reality by majority vote.  

I mean the above quite literally.  Setting $\pi$ equal to three is actually a good approximation to this irrational number.  Whenever I am doing a back of the envelope calculation, I usually do set $\pi$ equal to three.  Whenever I get out my calculator, I use a higher rational approximation: $\pi = 3.141592654$ is visible on my calculatorÕs screen, and there are several more digits that are in the calculatorÕs memory, but hidden from me.  If I need even higher accuracy, I can program {\it Mathematica} to generate a rational approximation good to several hundred decimal places.  But I will always approximate the irrational $\pi$ with a rational number.

By basing probability on the Kolmogorov Axioms, mathematicians are necessarily restricting probability to that tiny class of problems in which only the frequency appears.  For Laplace and Gauss and modern Bayesians, it makes perfect sense to speak of the probability of a single event.  It makes no sense to speak of the probability of a single event if a probability is a frequency.  .  Kolmogorov mathematicians are insisting that $\pi$ is exactly three.  The number $\pi$ is approximately three.  A probability is approximately a frequency, in the sense that in the limit that the number of measurements approaches infinity, the frequency approaches the probability.  Physically, no one will ever carry out a measurement of a probability if it were indeed a frequency, because it is impossible to carry out an infinite number of measurements.  Physically, the quantum mechanical probability is not a frequency.  As we shall see below, it is an expression of the indistinguishability of the analogues of the humans doing the experiment in the universes of the multiverse.
 
The frequency interpretation would mean either that there really is an irreducible element of chance in reality, completely undetermined by any laws of physics --- God really does play dice with the universe --- or there is a deeper theory than quantum mechanics, which is deterministic.  As stated earlier, in their heart of hearts, even physicists who are ignorant of the Laplacean theory of probability, incline to the latter view, believing with Einstein that the idea of ultimate indeterminism is not only nonsense,but objectionable nonsense.  

Just so.  The course of atoms, and their spins, in the multiverse is exactly and completely determined by the deterministic wave equation.  Further, the fact that it is the square of the modulus of the wave function that gives the best estimate of the probability density --- in the sense of a numerical value of human ignorance --- is due to the deterministic nature of the wave equation itself.

To see this, let us consider the Schr\"odinger equation for spinless particles with potential 

\begin{equation}
i\hbar\frac{\partial\psi}{ \partial t} = -\frac{\hbar^2}{
2m}\nabla^2\psi + V(\vec x)\psi
 \label{eq:QM}
\end{equation}	

Bohm \cite{Bohm52a}, \cite{Bohm52b} and Landau (\cite{Landau77}, p. 51--52) pointed out that the substitution 

\begin{equation}
\psi = {\cal R}\exp(i{\varphi}/h)
\label{eq:Pilotwave}
\end{equation}	

\noindent
for two real functions ${\cal R} = {\cal
R}(\vec x,t)$ and $\varphi = \varphi(\vec x,t)$ yields the two equations

 \begin{equation}
 \frac{\partial \varphi}{\partial t} = - \frac{(\vec\nabla
\varphi)^2}{ 2m} - V  +\left(\frac{\hbar^2}{2m}\right) \frac{\nabla^2{\cal R}}{{\cal R}}
\label{eq:Bohm2}
\end{equation}

\begin{equation}
\frac{\partial{\cal R}^2}{\partial t} + \vec\nabla\cdot
\left({\cal R}^2\frac{{\vec\nabla} \varphi}{m}\right)=0
\label{eq:Bohm1}
\end{equation}

Equation (\ref{eq:Bohm2}) is recognized as the classical Hamilton-Jacobi
equation for a single particle moving in the modied potential \cite{Jammer74}:

 \begin{equation}
U = V  -\left(\frac{\hbar^2}{
2m}\right)\frac{\nabla^2{\cal R}}{ {\cal R}}
 \label{eq:quanpot}
\end{equation}

This modified potential is required by global determinism: without it, a general nonsingular potential $V(\vec x)$ will cause caustic singularities to develop in the wave $\varphi$ after a finite time.  As an example of this, consider a spherically symmetric attractive potential that drops off as $r^{-1}$ far from the center of the potential and a wave which is plane far from this center.  The normals to the wave fronts (the tangents to the particle trajectories) will be focussed a finite distance behind the center, and this focal point will be the cusp singularity where the normal to the wave ceases to be defined.  With the modified potential (\ref{eq:quanpot}) this will never happen, because the added term forces the trajectories apart (Schr\"odinger's equation is linear, and hence cannot develop caustics).  The universes of the multiverse governed by the modified classical Hamilton-Jacobi equation are conserved.

The Hamilton-Jacobi equation was recognized both in the nineteenth century and today as the most powerful mathematical expression of classical mechanics.  But it is clear that the Hamilton-Jacobi equation is a multiverse expression of classical mechanics.  In the nineteenth century and often even today, this multverse nature was not taken seriously --- both then and now physicists have difficulty taking the fundamental equations of physics seriously; they cannot accept that their equations may be in one-to-one correspondence with reality --- and so only one trajectory of the Hamiltonian-Jacobi equation was believed to actually exist.  But the other worlds of the multiverse really do exist even in classical mechanics: it is the collision of the worlds that yield the caustics.  Both classical mechanics and quantum mechanics are multiverse theories. Quantum mechanics is nothing but classical mechanics made globally deterministic by the addition of a term to the potential.

Equation \ref{eq:Bohm1} is the conservation equation for the universes, and it is expressed in standard form for a conservation equation, which therefore allows us to recognize that ${\cal R}^2$ is the density of universes.  Thus the total number of what I shall term ``effectively distinguishable'' universes is the space integral of ${\cal R}^2$, and this integral may be infinite.  The total number of universes is necessarily uncountably infinite if ${\cal R}^2$ varies continuously with space and time, as it does in the Schr\"odinger equation.  However, the expression ${\cal R}^2d^3x$ can still be considered as describing a single universe if $d^3x$ is an infinitesimal volume element.   Astrophysicists (see any elementary astrophysics text book, e.g., \cite{CarrollOstlie}, p. 82), correctly term $L_\lambda d\lambda = d\lambda(4\pi A\hbar c^2)/(\lambda^5[e^{2\pi\hbar c/\lambda kT)} - 1])$ the ``monochromatic luminosity'' of a star --- i.e., literally, the luminosity of a star at a {\it single} wavelength --- with surface area $A$, with $d\lambda$ an infinitesimal bandwidth, even though there are literally an uncountable infinity of wavelengths present in any finite bandwidth.  We see that Schr\"odinger's equation does not require the integral of ${\cal R}^2$ to be finite, and there will be many cases of physical interest in which it is not.  The plane waves are one important and indispensable  example, and physicists use various delta function normalizations in this case.  An infinite  integral of ${\cal R}^2$ for the wave function of the multiverse has been shown \cite{Tipler2005} to provide a natural and purely kinematic explanation for the observed flatness of the universe.

However, in most cases of physical interest, the integral of ${\cal R}^2$ will be finite, and if we pose questions that involve the ratio of the number of ``effectively distinguishable'' worlds to the total number of ``effectively distinguishable'' worlds, it is convenient to normalize the spatial integral of ${\cal R}^2$ to be 1.

With this normalization, ${\cal R}^2d^3X$ is then the ratio of the number of ``effectively distinguishable'' universes in the region $d^3x$ to the total number of universes.  In the case of spin up and spin down, there are only two possible universes, and so the general rule for densities requires us to have the squares of the coefficients of the two spin states be the total number of effectively distinguishable -- in this case obviously distinguishable --- states.  Normalizing to 1 gives the ratio of the number of the two spin states to the total number of states.

Let us consider a two-electron state in more detail.  Specifically, consider two
spin 1/2 particles, with the two-particle system being in the rotationally invariant singlet state with zero total spin angular momentum.  Thus, if we decide to measure
the particle spins in the up-down direction, we would
write the wave function of such a state as

\begin{equation}
|\Psi> = \frac{|\uparrow >_1|\downarrow>_2 - 
 |\downarrow >_1|\uparrow>_2}{\sqrt2}
\label{eq:HD1}
\end{equation}

\noindent
where the direction of the arrow denotes the direction of spin, and the subscript identifies the particle.  To see that in quantum mechanics, as in classical Bayesian probability theory, probabilities are a measure of human ignorance rather than an intrinsic property of nature, we must apply quantum mechanics to both the observer and the observed, that is, we must assume once again the validity of the Many-Worlds Interpretation (MWI) of quantum mechanics.  All other interpretations either assume that quantum mechanics does not apply to the observer (the standard Copenhagen Interpretation) or are mathematically equivalent to the MWI.   Probabilities are a precise measure of human ignorance of the alternative versions of themselves in the various worlds of the multiverse.

The importance of applying quantum mechanics to the observer is immediately seen in the fact that the two spin 1/2 particle singlet state can also be written as

\begin{equation}
|\Psi> = \frac{|\leftarrow >_1|\rightarrow>_2 - 
 |\rightarrow >_1|\leftarrow>_2}{\sqrt2}
\label{eq:HD2}
\end{equation}

Which is correct?  The answer, of course, is that both are correct.  But equation (\ref{eq:HD1}) is more appropriate if the measuring apparatus is set to measure the spin as spin up or down --- in the vertical direction --- and equation (\ref{eq:HD2}) is more appropriate if the measuring apparatus is set to measure the spin as left or right --- in the horizontal direction.

To fix ideas, let the two spin 1/2 particles be electrons, and let the measuring apparatus be the Stern-Gerlach apparatus.  Then the orientation of the magnets, vertical or horizontal, determines whether the spin will be measured spin vertical or horizontal respectively.  As Bryce DeWitt has pointed out, the spin measurement is recorded in the motion of the center of mass of the atom to which the electron is bound; it can therefore record spin of any orientation.  If the electron spin is spin up, the atom moves up after passing through the magnetic field of the Stern-Gerlach apparatus.  The motion of the atomic center of mass can be recorded by more macroscopic objects, say a laboratory record book, but this will not alter the basic analysis.  Both the atomic center of mass and the record book are universally regarded by physicists as classical objects in this experiment, so they are.  All objects, from the Many-Worlds point of view are all equally quantum mechanical, and in certain circumstances the effects of the other other versions of the object can be ignored.  Let $M_i(...)$ be the state of the atomic center of mass before the passage through the magnetic field (or for a record book, before the spin is recorded) of the $i$th electron.  To measure the spin of two electrons, we will need two Stern-Gerlach apparatuses, whose interaction with the $i$th electron we will denote by the linear unitary operator ${\cal U}_i$, whose effect is completely described by it effects on the basis states.  For two vertically oriented Stern-Gerlach apparatuses, it is

\begin{eqnarray}
{\cal U}_1M_1(...)|\uparrow>_1 = M_1(\uparrow)|\uparrow>_1\nonumber\\
{\cal U}_1M_1(...)|\downarrow>_1 = M_1(\downarrow)|\downarrow>_1\label{eq:Uoperator1}\\
{\cal U}_2M_2(...)|\uparrow>_2 = M_2(\uparrow)|\uparrow>_2\nonumber\\
{\cal U}_2M_2(...)|\downarrow>_2 = M_2(\downarrow)|\downarrow>_2\label{eq:Uoperator2}
\end{eqnarray}

\noindent
where the combined state of the electron spin and the atomic center of mass (or the electron spin and the notebook) is represented by the product of the wave functions of the two objects.  In particular, if particle 1 is in an eigenstate of spin up, and particle 2 is in an eigenstate of spin down, then the effect of the ${\cal U}_i$'s together is

\begin{equation}
{\cal U}_2{\cal U}_1M_1(...)M_2(...)|\uparrow>_1
|\downarrow>_2 
 = M_1(\uparrow)M_2(\downarrow)
|\uparrow>_1|\downarrow>_2
\label{eq:U1U2operator}
\end{equation}

Then the effect of the two vertically oriented Stern-Gerlach apparatuses, both measuring the spins of the electrons is

\begin{eqnarray}
{\cal U}_2{\cal U}_1M_2(...)M_1(...)\left[{{|\uparrow>_1|\downarrow>_2 -
|\downarrow>_1|\uparrow>_2}\over \sqrt2}\right] =\nonumber\\
{\cal U}_2M_2(...)\left[{M_1(\uparrow)|\uparrow>_1|\downarrow>_2\over\sqrt2} -   {M_1(\downarrow)|\downarrow>_1| \uparrow>_2\over\sqrt2} \right]=\nonumber\\
{M_2(\downarrow)M_1(\uparrow)|\uparrow>_1|\downarrow>_2\over\sqrt2} - {M_2(\uparrow)
M_1(\downarrow) |\downarrow>_1|\uparrow>_2\over\sqrt2}\,\,\,\,\,\,\,\label{eq:U1U2M1M2}
\end{eqnarray}

The last line of (\ref{eq:U1U2M1M2}) is in the older literature on the Many-Worlds said to indicate that the universe has ``split'' into two universes, in one of which electron 1 has spin up and electron 2 has spin down, and in the other universe, electron 1 has spin down, and electron 2 has spin up.  However, since in the state space of quantum mechanics, multiplication distributes over addition:

\begin{eqnarray}
M_2(...)M_1(...)\left[\frac{{|\uparrow>_1|\downarrow>_2 -
|\downarrow>_1|\uparrow>_2}}{\sqrt2}\right] =\nonumber\\
\frac{M_2)(...)M_1(...)\left[|\uparrow>_1|\downarrow>_2\right]}{\sqrt2}\,\,\, -\nonumber\\
\frac{M_2)(...)M_1(...)\left[|\downarrow>_1|\uparrow>_2\right]}{\sqrt2}
\label{eq:distribute}
\end{eqnarray}

\noindent
equation (\ref{eq:U1U2M1M2}) could have been written:

\begin{eqnarray}
{\cal U}_2{\cal U}_1M_2(...)M_1(...)\left[\frac{|\uparrow>_1|\downarrow>_2 -
|\downarrow>_1|\uparrow>_2}{\sqrt2}\right] =\nonumber\\
\frac{{\cal U}_2{\cal U}_1M_2)(...)M_1(...)\left[|\uparrow>_1|\downarrow>_2\right]}{\sqrt2}\,\,\, -\nonumber\\
\frac{{\cal U}_2{\cal U}_1M_2)(...)M_1(...)\left[|\downarrow>_1|\uparrow>_2\right]}{\sqrt2}\nonumber\\
\frac{M_2(\downarrow)M_1(\uparrow)|\uparrow>_1|\downarrow>_2}{\sqrt2} -
\frac{M_2(\uparrow)M_1(\downarrow) |\downarrow>_1|\uparrow>_2}{\sqrt2}\,\,\,\,\,\,\,\label{eq:duplicate}
\end{eqnarray}

The only difference between (\ref{eq:U1U2M1M2}) and (\ref{eq:duplicate}) is the middle line.  But the middle line in (\ref{eq:duplicate}) indicates that the universe is ``split'' {\it before} the measurement is carried out, or rather, that the multiverse consists of the distinct universes before the measurement is carried out, the measurement having the effect of causing the universes to differentiate.  Since the two descriptions are mathematically equivalent, they are the same physically, and I shall use the two languages of ``splitting universes`` and ``differentiating universes'' interchangeably.  But how the universes split or differentiate, into universes in which the electron spin is spin up and down, or left or right, is determined by the specific interaction we chose to use.  I shall in what follows not explicitly show the measurement states, but they are physically present, and they, together with the measurement interaction, determine the state of the multiverse, and the probabilities we assign to the outcome of the measurements.

To show how probability comes about by measurements splitting (differentiating) the universe into distinct worlds, I follow \cite{Green90} and write the singlet state \ref{eq:HD1} with respect to a basis in a more general direction ${\bf\hat n}_1$ as

\begin{equation}
|\Psi> = (1/\sqrt2)(|{\bf\hat n}_1, \uparrow>_1 |{\bf\hat
n}_1,\downarrow>_2 - |{\bf\hat n}_1, \downarrow>_1 |{\bf\hat
n}_1, \uparrow>_2) \label{eq:psisinglet}
\end{equation}

Let another direction ${\bf\hat n}_2$ be the polar
axis, with $\theta$ the polar angle of ${\bf\hat n}_1$
relative to ${\bf\hat n}_2$.  Without loss of generality,
we can choose the other coordinates so that the
azimuthal angle of ${\bf\hat n}_1$ is zero.  Standard
rotation operators for spinor states then give\cite{Green90}

\begin{eqnarray*}
|{\bf\hat n}_1,\uparrow>_2 = (\cos\theta/2) |{\bf\hat n}_2, \uparrow>_2 \;+\; (\sin\theta/2)|{\bf\hat n}_2,\downarrow>_2\\
|{\bf\hat n}_1,\downarrow>_2 = -\; (\sin\theta/2)|{\bf\hat n}_2, \uparrow>_2 \;+
\;(\cos\theta/2)|{\bf\hat n}_2, \downarrow>_2
\end{eqnarray*}

\noindent
which yields

\begin{eqnarray}
|\Psi> = (1/\sqrt2)[\; - \;(\sin\theta/2) |{\bf\hat
n}_1,\uparrow>_1|{\bf\hat n}_2,\uparrow>_2\nonumber\\
\;+\;(\cos\theta/2) |{\bf\hat
n}_1,\uparrow>_1|{\bf\hat n}_2,\downarrow>_2\nonumber\\
 -\; (\cos\theta/2) |{\bf\hat
n}_1,\downarrow>_1|{\bf\hat n}_2,\uparrow>_2\nonumber\\
\;-\;(\sin\theta/2) |{\bf\hat
n}_1,\downarrow>_1|{\bf\hat n}_2,\downarrow>_2]
\label{eq:psisintheta}
\end{eqnarray}

In other words, if two different devices measure the spins of the two electrons
in two distinct arbitrary directions, there will be a split into {\it
four} worlds, one for each possible permutation of the
electron spins.  Just as in the case with ${\bf\hat n}_1
= {\bf\hat n}_2$, normalization of the devices on
eigenstates plus linearity forces the devices to split
into all of these four worlds, which are the only
possible worlds, since each observer must measure the
electron to have spin $+1$ or $-1$.

The fact that the splits are determined by the nature of the measurement apparatus is the key to deriving the Born Interpretation (BI) wherein the squares of the coefficients in (\ref{eq:psisintheta}) are the ``probabilities'' of an observed occurrence of the four respective outcomes in  (\ref{eq:psisintheta}).  Note that all two or four outcomes actually happen: the sums in (\ref{eq:psisintheta}) (or (\ref{eq:HD1}) are in 1 to 1 correspondence with {\it real} universes.  Since the observers are unaware of the other versions of themselves after a measurement, ignoring the existence of the other versions necessarily means a loss of information available to one observer, and it is this loss of information that results in probabilities.  The information is still in the collection of observers --- time evolution is unitary --- but it is now divided between the four versions, who are now mutually incommunicado.

Consider a measurement of (\ref{eq:HD1}) or (\ref{eq:psisintheta}) with $\theta = \pi/2$.  In either case, the initial state of the observer is the same, and there is no way even in principle of distinguishing the two or four final states of (\ref{eq:HD1}) or (\ref{eq:psisintheta}) respectively. Since there is no difference between the initial state observer, there is no difference in the terms of the expression except for the labels I have given them, and the labels can be interchanged leaving the physics invariant.  This interchange of labels forms a group, and shows that the probabilities assigned to each state must be the same.  This the transformation group argument for assigning a probability distribution is originally due to Henri Poincar\'e \cite{Poincare}; see (\cite{Jaynes03}, \cite{Sivia96}) for a modern discussion. Thus the invariance of the physics under the relabeling of the states yields the ``Principle of Indifference'': we must assign equal probabilities to each of two or four states respectively, and  so the probabilities must be $1/2$ or $1/4$ respectively.  These are seen to be the relative numbers of distinguishable universes in these states.  In summary, it is the indistinguishability of the initial state of the observer in all two or four final states that forces us to equate the probabilities with the relative number of distinguishable universes in the final state.  The same argument gives the same equation of the probability of the general orientation state in (\ref{eq:psisintheta}) with arbitrary $\theta$ with the squares of the coefficients of the states in (\ref{eq:psisintheta}) with the relative number of effectively distinguishable universes in the final states.

Notice that this does not give the Born Interpretation in the usual sense of ``probabilities mean relative frequencies as the number of observations approaches infinity."  In Laplacean probability theory, the relative frequency is a parameter to be estimated from a probability, not a probability itself (see \cite{Jaynes03}, \cite{Sivia96} for a detailed discussion of this point).  However, the most probable value of the relative frequency has been shown (\cite{Jaynes03},pp. 336--339, 367-368, 393-394, 576--578; \cite{Sivia96}, pp.106--110) to  be equal in classical physics to the probability (in the Laplacean sense) that the event will occur.

A proof in quantum physics that in the limit of a very large number of trials, the measured relative frequencies will approach the probabilities --- the measure of human ignorance of the other universes of the multiverse --- proceeds as follows.  The proof depends crucially on the indistinguishability of the initial states of the observer, and on the actual existence of the many worlds.  For simplicity I shall assume that the spins of a series of electrons are measured, and that the spins of all the measured electrons are spin up before the measurement.  I shall also assume that the measuring apparatus is at an arbitrary angle $\theta$ with respect to the vertical in all the universes.  In this case the Laplacean probabilities for measuring spin up along the axis of the apparatus is $p_{\uparrow,\,\theta} = \cos^2(\theta/2)\equiv p$ and for  measuring spin as anti-aligned with the axis is $p_{\downarrow,\,\theta} = \sin^2(\theta/2)\equiv q$, respectively, for $0\leq\theta\leq\pi/2$.  The probability ${\rm prob}(r|N)$ that an observer in a particular universe will, after $N$ measurements of N different electrons but with all in the spin up state, see the electron as having spin aligned with the apparatus $r$ times, is

\begin{eqnarray}
{\rm prob}(r|N) = \sum_k\,{\rm prob}(r, S_k|N)\nonumber\\
=  \sum_k\,{\rm prob}(r| S_k,\, N)\times {\rm prob}(S_k|N)
\label{eq:prob(r|n)}
\end{eqnarray}

\noindent
where the summation is over all the $2^N$ sequences of outcomes $S_k$, each of which actually occurs in some universe of the multiverse, after $N$ measurements in each of these now $2^N$ distinct universes.  The first term in the second line of (\ref{eq:prob(r|n)}) will equal one if $S_k$ records exactly $r$ measurements of the spin in the $\theta$ direction, and will be zero otherwise.  Since the $N$ electrons are independent, the probability of getting any particular sequence $S_k$ depends only on the number of electrons with spins measured to be in the $\theta$ direction, and on the number with spins measured in the opposite direction.  In particular, since the only sequences that contribute to (\ref{eq:prob(r|n)}) are those with $r$ spins measured to be in the $\theta$ direction and those with $N-r$ spins to be in the opposite direction, we have

 \begin{equation}
{\rm prob}(S_k|N) = p^r q^{N-r}
 \label{eq:prob(S_k|N)}
\end{equation}

However, the order in which the $r$ aligned spins and the $N-r$ anti-aligned spins are obtained are irrelevant, so the number of times (\ref{eq:prob(S_k|N)}) appears in the sum (\ref{eq:prob(r|n)}) will be $C^N_r$, the number of combinations.  Thus the sum (\ref{eq:prob(r|n)}) is

 \begin{equation}
{\rm prob}(r\,|\, N) = \frac{N!}{r!(N-r)!}p^r q^{n_r}
 \label{eq:prob(r|N}
\end{equation}	

The relative number of universes in which we would expect to measure aligned spin $r$ times ---that is to say, the expected value of the frequency with which we would measure the electron spin to be aligned with the axis of the measuring apparatus --- is
\begin{equation}
\langle f \rangle = \Big\langle \frac{r}{N}\Big\rangle = \sum_{r=0}^N \Big(\frac{r}{N}\Big){\rm prob}(r\,|\, N)\nonumber
\end{equation}

\begin{equation}
=  \sum_{r=1}^N \frac{(N-1)!}{(r-1)!(N-r)!}p^r q^{N-r} = p(p+q)^{N-1} = p
\label{eq:freq}
\end{equation}

\noindent
where the lower limit has been replaced by one, since the value of the $r=0$ term is zero.

The sum in the second line of (\ref{eq:freq}) has been evaluated by differentiating the generating function of the binomial series $\sum_{r=0}^NC^N_r p^rq^{N-r} = (p+q)^N$.  That is, we have $\langle r^m \rangle = (p[d/dp])^m(p+q)^N$, where $q$ is regarded as a constant in the differentiation, setting $p+q = 1$ at the end.  This trick also allows us to show that the variance of the difference between the frequency $f = r/N$ and the probability $p$ vanishes as $N \rightarrow \infty$, since we have

 \begin{equation}
\Big\langle\left( \frac{r}{N} - p\right)^2\Big\rangle = \frac{pq}{N}
 \label{eq:varf}
\end{equation}	
 
\noindent
In fact, all moments of the difference between $f$ and $p$ vanish as $N \rightarrow \infty$, since the generating function gives

 \begin{equation}
\Big\langle\left( \frac{r}{N} - p\right)^{m}\Big\rangle \sim \frac{1}{N} + {\rm higher\, order\,terms\, in }\,\frac{1}{N}
 \label{eq:momentf}
\end{equation}	

So we have

 \begin{equation}
\lim_{N\to\infty}\left( \frac{r}{N}\right) = p
 \label{eq:limtf}
\end{equation}	

\noindent
in the sense that all the moments vanish as $1/N$ as $N \rightarrow \infty$.  This law of large numbers explains why it has been possible to believe, incorrectly, that probabilities are frequencies.   Not so, as Laplace emphasized over two hundred years ago.  It is, instead, that the quantum property of indistinguishability, applied to the observers, forces the measured frequencies to approach the probabilities.

Physicists are used to the notion of indistinguishability applied to atomic and subatomic systems.  Interchanging two atoms in their ground state, or two electrons in the atom, or two photons with the same energy, momentum and polarization, means that nothing whatsoever has happened.  Applying this indistinguishability yields Fermi-Dirac statistics for fermions, and Boson-Einstein statistics for bosons.  It is well-known that Willard J. Gibbs introduced this indistinguishability into statistical mechanics in 1875 to resolve the Gibbs Paradox.  It is less well-known that James Clerk Maxwell made use of indistinguishability in 1878 to show it implied the quantization of energy in certain macroscopic physical systems (\cite{Tipler1994}.  I shall follow Maxwell and assume that this quantum property of indistinguishability applies to the macroscopic systems which we are pleased to call ``observers,'' as well as to atoms and molecules.  This will allow us to deduce the numerical probabilities of quantum mechanics as an exact measure of human ignorance, as a numerical expression of the human ignorance of the Many-Worlds.

It cannot be emphasized too strongly that a ``probability'' cannot be an objective feature of reality, but instead a ``probability'' is a numerical expression of human ignorance of the actual state of affairs.  Consider the canonical example of probability theory, the probability that a fair coin will come up heads.  To say that it is 1/2 independently of the way the coin is tossed is actually an assertion that the coin is not subject to the laws of mechanics, specifically the laws of conservation of angular momentum.  If a ``fair coin'' is defined to be one whose geometric center is very close to its center of gravity, then a skilled coin tosser can use conservation of angular momentum to arrange the coin to land heads almost all the time, and do so in such a way that an untrained observer cannot tell this is due to the particular way the coin is tossed.  However, if by ``probability'' we mean ``measure of human ignorance'' in this case, either ignorance on the part of the observer who is unaware that he is dealing with a skilled coin tosser, or ignorance in the case of the coin tosser, who is unaware of how to toss the coin so that it almost always come up heads, then the probability is indeed 1/2, because in the former case, the ignorant human does not know any reason why heads should be preferred over tails, and in the latter case, a knowledgeable observer has no reason to believe the ignorant coin tosser will systematically select a particular toss technique that will favor heads over tails.  Both beliefs can be wrong, and probabilities reassigned after more knowledge becomes available.    But we cannot improve our knowledge in the quantum mechanical case.  Quantum indistinguishability and our ignorance of the other universes preclude an increase in knowledge.

We must remember, however, that David Deutsch has shown \cite{Deutsch86} that a non-human mentality, based on a quantum computer, can be aware of more than one universe.  Such a mentality would not necessarily observe the Born Interpretation frequencies.  Since this shows that the Born frequencies need not hold for some (nonhuman) observer, it is possible that humans could devise an experiment that would deviate from the Born Interpretation.

Notice that the above derivation of the measured frequencies require the actual existence of the other universes of the multiverse.  All of the sequences $S_k$ really exist.  The fact that the measured frequencies approach the probabilities requires that the indistinguishable versions of the physicist carry out the measurements simultaneously.  So an observation of the approach of the frequencies to the probabilities is actually an observation of the effect of the simultaneous action in the multiverse of the analogues of the human observer.

Similarly, in a quantum computer, a device first described by David Deutsch(\cite{Deutsch85}), the human analogues of the person setting up the program to be  run on the quantum computer are cooperating across the multiverse to compute an answer which they will all share.  So physicists also are in a sense interacting with other versions of themselves --- are ``seeing'' other versions of themselves whenever they run a computation on a quantum computer.

Using an incorrect probability theory has prevented physicists from realizing that they have actually directly observed the effects of the other versions of themselves.

The historians of physics Max Jammer (\cite{Jammer66}, pp. 166--167), Paul Forman \cite{Forman71}, and Stephan G. Brush \cite{Brush78} have presented strong and convincing evidence that the Born Interpretation in the sense that probabilities are frequencies and not a mere mathematical expression of human ignorance of the underlying determinism of physical reality, did not originate in physics at all.  Born and all the others later associated with the idea that quantum mechanics, and hence physical reality, were fundamentally non-deterministic had defended this idea years before Schr\"odinger published his equation.  Their belief in indeterminism was a reflection, not of physical concepts, or experiments, but of the dominant political and philosophical ideas in the Weimar Republic.  It is long past time for physicists to accept the Laplacean theory of probability, a theory developed by great physicists for physicists, and based on observations and the laws of physics.  Physicists must reject philosophical and politically-based ideas like indeterminism and the frequency interpretation of probability.  Philosophy and politics have no place in physics.

This is not to say the CI cannot be given a valid meaning.  Follow Bohr and require that the measurement basis must be classical, but let ``classical'' mean that the wave function of the ``classical'' entity has zero quantum potential.  For such $\Psi$, Schr\"odinger's equation becomes identical to the Hamilton-Jacobi equation, which is the expression of the MWI in classical mechanics.  If $V(x) = 0$, then plane waves are the most classical state, and not the minimal uncertainty wavepacket, since for plane waves, with constant modulus, the quantum potential vanishes.

An even more powerful argument for the MWI is the true origin of the BI, which I derived above.  The main difficultly that many physicists have had with the MWI is the required existence of the analogues of themselves.  But every time physicists measure a frequency, and verify the quantum expectation value in the Bell inequality, they are actually seeing the effect of the analogues of themselves making the same measurements of the electron spin$^{27}$.  The language of the frequency interpretation of probability has prevented physicists from seeing what is actually happening.  It has prevented physicists from seeing the other universes of the multiverse.

Have you ever seen the Earth rotate on its axis?  I have.  I see it every day, when I see the Earth's rotation expose the unmoving Sun at dawn, and cover the unmoving Sun from my view at dusk.  Common language, however, says that the Sun sets, and also rises.  So everyone believed until Copernicus and Galileo taught not only physicists, but almost everyone, to see Nature through the laws of physics.  It is time to see the measurements of the electron spin frequencies through the laws of quantum mechanics, which applies not only to electrons, but also to the physicists who measure these spins.

\medskip
There are none so blind as those who will not see.

\medskip
{\it Acknowledgments}  I am grateful to David Deutsch, Maurice Dupre, and Paul Forman for helpful comments.


\begin{thebibliography}{99}

\bibitem{Jaynes03}E. ~T. ~Jaynes {\it Probability Theory: The Logic of Science} (Cambridge Univ. Press, Cambridge, 2003).

\bibitem{Sivia96}D.~S.~Sivia {\it Data Analysis: A Bayesian Tutorial} (Clarendon Press, Oxford, 1996). 

 \bibitem{Darwin1868} C. ~Darwin {\it The Variation of Animals and Plants Under Domestication, first edition} (John Murray, London, 1868).
 
 \bibitem{Darwin1860}C. ~Darwin {\it On the Origin of Species by Means of Natural Selection, or the Preservation of Favoured Races in the Struggle for Life}. (John Murray, London, 1860).

\bibitem{Green90}D.~M.~Greenberger {\em et al}, Am. J. Phys. {\bf 58}, 1131 (1990). 

\bibitem{Bohm52a}David Bohm ``Suggested Interpretation of the Quantum Theory in Terms of Hidden Variables: 1''  {\it Phys Rev} {\bf 85} (1952), 166--179.

\bibitem{Bohm52b} David Bohm  ``Suggested Interpretation of the Quantum Theory in Terms of Hidden Variables: 2''  {\it Phys Rev} {\bf 85} (1952) 180--193.

\bibitem{Landau77}Lev D Landau and E M Lifshitz 1977 {\it QuantumMechanics: Non-relativistic Theory, 3rd edition} (Oxford: Pergamon Press).

\bibitem{Jammer74} Max Jammer 1974 {\it The Philosophy of Quantum Mechanics} (New York: Wiley).

\bibitem{EinsteinPlanck}Albert Einstein and Max Planck {\it Where is Science Going?} translated and edited by James Murphy.  (New York: Norton, 1932), pp. 201--202.

\bibitem{Laplace} Pierre-Simon de Laplace {\it Philosophical Essay on Probabilities} (Dover, New York, 1996). Note that the word ``philosophical'' to Laplace meant ``natural philosophy,'' i.e., physics.
\bibitem{Schroedinger}E. Schr\"odinger, Proc. Roy. Irish Acad. {\bf 51A} (1947), 51--61, largely re-discovered the Laplacean probability theory shortly before he, independently of Everett, discovered the MWI.  Unfortunately, Schr\"odinger did not go on to connect his discoveries, and thereby derive the Born Interpretation from determinism in quantum mechanics.

\bibitem{Poincare} Henri Poincar\'e {\it Calcul des Probabilit\'es}, second edition, (Gauthier-Villars, Paris, 1912).

\bibitem{Tipler1994} Frank J. Tipler {\it The Physics of Immortality} (Doubleday, New York 1994), see pp. 230--231 for a discussion of this virtually unknown achievement of Maxwell.  Maxwell' s original paper was an article for {\it Encyclopaedia Britannica}.  This article, entitled ``Diffusion,'' was reprinted in {\it The Scientific Papers of James Clerk Maxwell}, Volume 2 (Cambridge University Press, Cambridge,1890) on pages 624--646, and the energy quantization argument is on pages 645--646.
\bibitem{Deutsch86} David Deutsch in {\it Quantum Concepts in Space and Time}, ed. R. Penrose and C.J. Isham (Clarendon Press, Oxford, 1986), pp. 215--225.

\bibitem{Deutsch85} David Deutsch {\it Proc. Roy. Soc. Lond.} {\bf A400} (1985), 77--117. 

\bibitem{Jammer66}Max Jammer {\it The Conceptual Development  of Quantum Mechanics} ( McGraw-Hill, NY, 1966).

\bibitem{Forman71}Paul Forman {\it Historical Studies in the Exact Sciences} {\bf 3} (1971) 1--115.

\bibitem{Brush78}S. G. Brush {\it Social Studies of Science} {\bf 10} (1978), 393--447.

\bibitem{CarrollOstlie} B. W. Carroll and D. A. Ostlie {\it An Introduction to Modern Astrophysics} (Addison-Wesley, New York, 1996).

\bibitem{Tipler2005}F. J. Tipler {\it Rep. Prog. Phys.} {\bf 68} (2005), 897--964.

\bibitem{Planck}Max Planck {\it The Philosophy of Physics} (Norton: New York, 1936).

\bibitem{Sommerfeld}Arnold Sommerfeld {\it Wave Mechanics} (Methuen: London, 1930).

\bibitem{Born}Max Born {\it Zeitschrift f\"r Physik} {\bf 38} (1926) 803; {\bf 40} (1928), 167.

\bibitem{Jordon}P. Jordon {\it Zeitschrift f\"r Physik} {\bf 37} (1926) 376; {\bf 40} (1927) 809; {\bf 41} (1927) 797; {\bf 44} (1927) 1.

\bibitem{Heisenberg}Werner Heisenberg  {\it Zeitschrift f\"r Physik} {\bf 40} (1927) 501; {\bf 43} (1927) 172.

\bibitem{Dirac}P.A.M. Dirac {\it Proc. Roy. Soc. A} {\bf 113} (1927) 621 

\bibitem{London}Fritz London {\it Zeitschrift f\"r Physik} {\bf 40} (1926) 193.

\bibitem{BHJ}M. Born, W. Heisenberg, and P. Jordon {\it Zeitschrift f\"r Physik} {\bf 35} (1926) 557.

\bibitem{Plato}J. von Plato {\it Creating Modern Probability: Its Mathematics, Physics, and Philosophy in Historical Perspective} (Cambridge: Cambridge University Press, 1994).

\bibitem{Howie}D. Howie {\it Interpreting Probability: Controversies and Development in the Early Twentieth Century} (Cambridge: Cambridge University Press, 2002).  I am grateful to Paul Forman for this reference.



\end{thebibliography}
\end{document}